\documentclass[11pt,preprint]{aastex}

\newcommand{\bfm}[1]{{\mbox{\boldmath $#1$}}}
\newcommand{\sbfm}[1]{{\mbox{\scriptsize\boldmath $#1$}}}
\newcommand{\ssbfm}[1]{{\mbox{\tiny\boldmath $#1$}}}

\begin{document}



\title{Phase Correlations in Non-Gaussian Fields}

\author{Takahiko Matsubara}
\affil{Department of Physics and Astrophysics, 
	Nagoya University, Chikusa, Nagoya 464-8602, Japan}

\email{taka@a.phys.nagoya-u.ac.jp}

\begin{abstract}
We present the general relationship between phase correlations and the
hierarchy of polyspectra in the Fourier space, and the new theoretical
understanding of the phase information is provided. Phase correlations
are related to the polyspectra only through the non-uniform
distributions of the phase sum $\theta_{\sbfm{k}_1} + \cdots +
\theta_{\sbfm{k}_N}$ with closed wave vectors, $\bfm{k}_1 + \cdots +
\bfm{k}_N = 0$. The exact relationship is given by the infinite
series, which one can truncate in a consistent manner. The method to
calculate the series to arbitrary order is explained, and the explicit
expression of the first-order approximation is given. A numerical
demonstration proves that the distribution of the phase sum is a
robust estimator and provides an alternative statistic to search for
the non-Gaussianity.
\end{abstract}



\section{Introduction}
\label{sec1}

Quantifying the cosmic fields, such as the density fields, velocity
fields, gravitational lensing fields, temperature fluctuations in the
cosmic microwave background, etc.~is undoubtedly crucial to study the
origin and dynamics of the structure in the universe. The structure of
these fields are believed to be emerged from primordial random
Gaussian perturbations, as most of the inflationary models naturally
predict nearly scale-invariant Gaussian fluctuations
\citep{gut82,sta82,haw82,bar83}.

Even if the primordial perturbations are random Gaussian, the
gravitationally nonlinear evolution produces non-Gaussianity in the
cosmic fields. Quantifying the non-Gaussianity is not trivial since it
depends on the full hierarchy of the higher-order correlation
functions in real space, or of the polyspectra in Fourier space. First
several members of such hierarchy can be observationally determined,
with which only partial information on non-Gaussianity is quantified.
Therefore, alternative statistics, such as the void probability
function \citep{whi79}, the genus statistic \citep{got86}, the
Minkowski functionals \citep{min03,mec94,sch97}, etc.~which contain
the information on the full hierarchy of higher-order statistics
should be useful.

Non-Gaussianity is frequently termed ``phase correlations''. This term
reflects the fact that the Fourier phases of a random Gaussian field
are randomly distributed without any correlation among different
modes. Therefore, phase correlations, if any, obviously characterize
the non-Gaussianity. However, what kind of phase correlations arise in
a given non-Gaussian field have been far from obvious. Investigations
along this line are quite limited in the literatures despite its
importance, apparently because of the lack of theoretical guidelines.
Most of the earlier work \citep{ryd91,sod92,ber98} only assess the
nonlinear evolution of phases in individual Fourier modes without
statistics. Phenomenological studies of $N$-body simulations have
revealed that the one-point phase distribution remains uniform even in
non-Gaussian fields \citep{sug91}, and that the phase difference
between neighboring Fourier modes is non-uniformly distributed
\citep{sch91,col00,chi01,chi02,wat03b}. However, the meaning of the
discovered phase correlations is obscure in those literatures.

Since the hierarchy of the higher-order statistics contains
statistically all information on the distribution \citep{ber92}, there
should be some connection between phase correlations and polyspectra,
which is the key to theoretically understand the phase correlations.
Examining a toy model, \citet{wat03a} realized the importance of the
phase sums with closed wave vectors in this connection, although they
have never derived the exact relations. In this Letter, the connection
in the general form is discovered for the first time. As a result, we
will have much better theoretical understanding of the phase
information than before.

\section{Phase correlations and polyspectra}
\label{sec2}

Although the real part ${\rm Re} f_\sbfm{k}$ and the imaginary part
${\rm Im} f_\sbfm{k}$ of the Fourier transform $f_\sbfm{k}$ of a
random field $f$ are naturally the independent variables, one can also
take their linear combinations $f_\sbfm{k} = {\rm Re} f_\sbfm{k} + i
{\rm Im} f_\sbfm{k}$ and $f_\sbfm{k}^* = {\rm Re} f_\sbfm{k} - i {\rm
Im} f_\sbfm{k}$ as another set of mutually independent variables. For
calculational advantages, we use the latter choice. In this Letter,
the reality of the random field $f$ is assumed since most of the
cosmic fields are real, although one can readily generalize the
following analysis to the complex fields. Because of the reality
condition, $f_\sbfm{k}^* = f_{-\sbfm{k}}$, $f_\sbfm{k}^*$'s are
actually not independent variables, and $f_\sbfm{k}$'s of all modes
$\bfm{k}$ are taken as independent variables. 

It is useful to define the normalized quantity $\alpha_\sbfm{k} \equiv
f_\sbfm{k}/\sqrt{P(k)}$, where $P(k) = \langle |f_\sbfm{k}|^2 \rangle$
is the power spectrum of the random field. The key technique to derive
the relation between phase correlations and polyspectra is given by
previous work \citep{mat95,mat03}: the joint probability function
$P(\{\alpha_\sbfm{k}\})$ of having particular set of $\alpha_\sbfm{k}$
is formally represented by
\begin{equation}
   {\cal P}(\{\alpha_\sbfm{k}\}) =
   \exp
   \left(
      \sum_{N=3}^\infty \frac{(-)^N}{N!}
      \sum_{\sbfm{k}_1,\ldots,\sbfm{k}_N}
      \left\langle
         \alpha_{\sbfm{k}_1}\cdots\alpha_{\sbfm{k}_N}
      \right\rangle_{\rm c}
      \frac{\partial^N}
         {\partial\alpha_{\sbfm{k}_1}\cdots\partial\alpha_{\sbfm{k}_N}}
   \right)
   {\cal P}_{\rm G}(\{\alpha_\sbfm{k}\}),
\label{eq01}
\end{equation}
where $\langle\cdots\rangle_{\rm c}$ indicates the cumulants, and
${\cal P}_{\rm G}(\{\alpha_\sbfm{k}\})$ is the multivariate Gaussian
distribution function of variables $\{\alpha_\sbfm{k}\}$. In the
present case, ${\cal P}_{\rm G}(\{\alpha_\sbfm{k}\}) \propto
\exp\left(-\frac12 \sum_\sbfm{k} \alpha_\sbfm{k}
\alpha_{-\sbfm{k}}\right)$, since the covariance matrix is $\langle
\alpha_\sbfm{k} \alpha_{\sbfm{k}'}\rangle = \delta^{\rm
K}_{\sbfm{k}+\sbfm{k}'}$, where the symbol $\delta^{\rm K}_{\sbfm{k}}$
is defined by $\delta^{\rm K}_{\sbfm{k}} = 1$ for $\bfm{k} = 0$ and
$\delta^{\rm K}_{\sbfm{k}} = 0$ for $\bfm{k} \ne 0$. The periodic
boundary condition with boxsize $V=L^3$ is assumed.

Since the polyspectra $P^{(N)}(\bfm{k}_1,\ldots,\bfm{k}_N)$ are
defined from the cumulants by
\begin{equation}
   \langle f_{\sbfm{k}_1}\cdots f_{\sbfm{k}_N} \rangle_{\rm c} =
   V^{1-N/2}
   \delta^{\rm K}_{\sbfm{k}_1 + \cdots + \sbfm{k}_N}
   P^{(N)}(\bfm{k}_1,\ldots,\bfm{k}_{N-1}),
\label{eq02}
\end{equation}
the above formula (\ref{eq01}) provides the relation between
polyspectra and joint distribution of the Fourier coefficients.
Expanding the exponential in equation (\ref{eq01}), each term in this
expansion consists of the products of polyspectra times derivatives of
${\cal P}_{\rm G}$. The derivatives of ${\cal P}_{\rm G}$ are given by
a simple polynomial of $\alpha_{\sbfm{k}}$'s times ${\cal P}_{\rm G}$.
The general term in the expansion has the form
\begin{equation}
   \sum_{\sbfm{k}'s}
   \delta^{\rm K}_{\sbfm{k}_1 + \sbfm{k}_2 + \cdots}
   \delta^{\rm K}_{\sbfm{k}'_1 + \sbfm{k}'_2 + \cdots} \cdots
   p^{(N)}(\bfm{k}_1,\bfm{k}_2,\ldots)
   p^{(N)}(\bfm{k}'_1,\bfm{k}'_2,\ldots) \cdots
   H_{\sbfm{k}_1 \sbfm{k}_2 \cdots \sbfm{k}'_1 \sbfm{k}'_2 \cdots}
   {\cal P}_{\rm G},
\label{eq03}
\end{equation}
with appropriate coefficients, where
\begin{equation}
   p^{(N)}(\bfm{k}_1,\bfm{k}_2,\ldots,\bfm{k}_{N-1}) =
   \frac{P^{(N)}(\bfm{k}_1,\bfm{k}_2,\ldots,\bfm{k}_{N-1})}
      {\sqrt{V^{N-2} P(k_1)P(k_2)\cdots P(k_{N-1})
       P\left(|\bfm{k}_1 + \cdots + \bfm{k}_{N-1}|\right)}},
\label{eq04}
\end{equation}
are the dimensionless, normalized polyspectra of $\alpha_{\sbfm{k}}$,
and
\begin{equation}
   H_{\sbfm{k}_1 \sbfm{k}_2 \cdots} =
   \frac{1}{{\cal P}_{\rm G}}
   \left(-\frac{\partial}{\partial\alpha_{\sbfm{k}_1}}\right)
   \left(-\frac{\partial}{\partial\alpha_{\sbfm{k}_2}}\right)
   \cdots
   {\cal P}_{\rm G},
\label{eq05}
\end{equation}
is a generalization of Hermite polynomials and is given by polynomials
of $\alpha_{\sbfm{k}}$'s and $\delta^{\rm K}_{\sbfm{k}}$'s. For
example,
\begin{equation}
   H_{\sbfm{k}_1 \sbfm{k}_2 \sbfm{k}_3} = 
   \alpha_{-\sbfm{k}_1}\alpha_{-\sbfm{k}_2}\alpha_{-\sbfm{k}_3} -
   \delta^{\rm K}_{\sbfm{k}_1 + \sbfm{k}_2}\alpha_{-\sbfm{k}_3} -
   \delta^{\rm K}_{\sbfm{k}_2 + \sbfm{k}_3}\alpha_{-\sbfm{k}_1} -
   \delta^{\rm K}_{\sbfm{k}_3 + \sbfm{k}_1}\alpha_{-\sbfm{k}_2},
\label{eq06}
\end{equation}
and so on. Thus, the joint probability function ${\cal
P}(\{\alpha_{\sbfm{k}}\})$ is represented by ${\cal P}_{\rm G}$ times
infinite sum of products by $\alpha_{\sbfm{k}}$'s, $\delta^{\rm
K}_{\sbfm{k}}$'s, and normalized polyspectra.

Next step is to transform the complex variable $\alpha_{\sbfm{k}}$
into the modulus $|\alpha_{\sbfm{k}}|$ and the phase
$\theta_{\sbfm{k}}$ by $\alpha_{\sbfm{k}} = |\alpha_{\sbfm{k}}|
e^{i\theta_{\ssbfm{k}}}$. This should be carefully done, since
$\alpha_{\sbfm{k}}$ is considered independent on
$\alpha_{\sbfm{k}}^*$. At this point, we restrict the wavenumber
$\bfm{k}$ in the upper half sphere (uhs), $k_z \ge 0$, and the degrees
of freedom in the lower half sphere is relabeled by the reality
relation, $\alpha_{\sbfm{k}} = \alpha_{-\sbfm{k}}^*$ for $k_z < 0$.
The mode $\bfm{k}=0$ is excluded which ensures zero mean of the
original field $f$. The term (\ref{eq03}) is accordingly relabeled,
resulting in the sum of the products of $\alpha_{\sbfm{k}}$'s,
$\alpha_{\sbfm{k}}^*$'s, $\delta^{\rm K}_{\sbfm{k}}$'s, where $\bfm{k}
\in {\rm uhs}$, and normalized polyspectra. With the above procedures,
one can express the ratio ${\cal P}(\{\alpha_{\sbfm{k}}\})/{\cal
P}_{\rm G}(\{\alpha_{\sbfm{k}}\})$ in terms of the normalized
polyspectra, the modulus $|\alpha_{\sbfm{k}}|$ and the phase
$\theta_{\sbfm{k}}$. The Jacobian of the transform from
$\alpha_{\sbfm{k}}$ to $(|\alpha_{\sbfm{k}}|, \theta_{\sbfm{k}})$ is
the same for the probability functions ${\cal
P}(\{\alpha_{\sbfm{k}}\})$ and ${\cal P}_{\rm
G}(\{\alpha_{\sbfm{k}}\})$. Therefore this meets our ends to relate
the phase correlations and polyspectra, which is a completely new
result.

Practically, one needs to truncate the infinite series by a consistent
manner. Fortunately, the non-Gaussianity generated by gravitationally
nonlinear evolution is known to approximately follow the hierarchical
model of the higher-order correlations, in which the polyspectra
$P^{(N)}$ have the order, $P^{(N)} \sim {\cal O}[P(k)^{N-1}]$
\citep[e.g.,][]{ber02}. This means $p^{(N)} \sim {\cal O}(\epsilon^{N-2})$,
where $\epsilon \sim \sqrt{P(k)/V}$. Therefore one can evaluate the
phase correlations in perturbative manner as long as the expansion
parameter $\epsilon$ is small. It is straightforward to perform the
above procedure to express ${\cal P}(\{|\alpha_{\sbfm{k}}|,
\theta_{\sbfm{k}}\})$ in terms of normalized polyspectra to arbitrary
order in $\epsilon$. In the lowest order approximation, only the
normalized bispectrum $p^{(3)}$ gives the term of order ${\cal
O}(\epsilon^1)$. The result is
\begin{eqnarray}
&& {\cal P}(\{|\alpha_{\sbfm{k}}|, \theta_{\sbfm{k}}\})
   \prod_{\sbfm{k}\in{\rm uhs}} d|\alpha_{\sbfm{k}}|
   d\theta_{\sbfm{k}}
\nonumber\\
&&\qquad = \left[1 +
   \sum_{\sbfm{k}_1,\sbfm{k}_2 \in {\rm uhs}}
   |\alpha_{\sbfm{k}_1}||\alpha_{\sbfm{k}_2}| |\alpha_{\sbfm{k}_1 +
   \sbfm{k}_2}|
   \cos\left( \theta_{\sbfm{k}_1} + \theta_{\sbfm{k}_2} -
   \theta_{\sbfm{k}_1 + \sbfm{k}_2} \right) p^{(3)}(\bfm{k}_1,
   \bfm{k}_2) \right]
\nonumber\\
&&\qquad\quad\times
   \prod_{\sbfm{k}\in{\rm uhs}} 2|\alpha_{\sbfm{k}}|
   e^{-|\alpha_{\ssbfm{k}}|^2}
   d|\alpha_{\sbfm{k}}|\frac{d\theta_{\sbfm{k}}}{2\pi}.
\label{eq07}
\end{eqnarray}
Higher-order terms can be similarly calculated, although they are
somehow tedious. For example, in the second-order approximation,
${\cal O}(\epsilon^2)$, there appears the square of the first-order
term, and terms like
\begin{equation}
   |\alpha_{\sbfm{k}_1}||\alpha_{\sbfm{k}_2}||\alpha_{\sbfm{k}_3}|
   |\alpha_{\sbfm{k}_1 + \sbfm{k}_2 \pm \sbfm{k}_3}|
   \cos\left(
      \theta_{\sbfm{k}_1} +
      \theta_{\sbfm{k}_2} \pm
      \theta_{\sbfm{k}_3} -
      \theta_{\sbfm{k}_1 + \sbfm{k}_2 \pm \sbfm{k}_3}
   \right)
\label{eq08}
\end{equation}
with appropriate normalized trispectrum or the product of normalized
bispectra multiplied, and other terms which do not depend on phases.

The phases always contribute to the probability distribution by the
combination of the form, $\cos(\theta_{\sbfm{k}_1} + \cdots +
\theta_{\sbfm{k}_N})$, with closed wavevectors: $\bfm{k}_1 + \cdots +
\bfm{k}_N = 0$. This is generally true because the phase dependence in
equation (\ref{eq03}) is the exponential of the sum of phases, and the
probability is the real number so that taking real parts gives the
cosine function. The reason that phase correlations exist only among
modes with closed wavevectors comes from the translational invariance.
In equations (\ref{eq07}) and (\ref{eq08}), wavenumbers are restricted
to the uhs so that the modes in the lower half sphere are relabeled by
$\theta_{\sbfm{k}} = - \theta_{-\sbfm{k}}$.

The moduli $|\alpha_{\sbfm{k}}|$'s are easily integrated in the
first-order approximation of equation (\ref{eq07}), resulting in
\begin{equation}
   {\cal P}(\{\theta_{\sbfm{k}}\}) \propto
   1 + \frac{\sqrt{\pi}}{2} \sum_{\sbfm{k}}^{\rm uhs}
   p^{(3)}(\bfm{k},\bfm{k})
   \cos(2\theta_{\sbfm{k}} - \theta_{2\sbfm{k}})
+
   \left(\frac{\sqrt{\pi}}{2}\right)^3
   \sum_{\sbfm{k}\ne \sbfm{k}'}^{\rm uhs}
   p^{(3)}(\bfm{k},\bfm{k}')
   \cos(\theta_{\sbfm{k}} + \theta_{\sbfm{k}'} -
      \theta_{\sbfm{k} + \sbfm{k}'}).
\label{eq09}
\end{equation}
The practically useful relations between phase correlations and the
bispectrum are obtained by further integrating some phases in equation
(\ref{eq09}). One obtains
\begin{eqnarray}
&&
   {\cal P}(\theta_\sbfm{k},\theta_{2\sbfm{k}}) \propto
   1 + \frac{\sqrt{\pi}}{2}
   p^{(3)}(\bfm{k},\bfm{k})
   \cos(2\theta_{\sbfm{k}} - \theta_{2\sbfm{k}}),
\label{eq10}\\
&& {\cal P}(\theta_\sbfm{k},\theta_{\sbfm{k}'},\theta_{\sbfm{k} +
   \sbfm{k}'}) \propto 1 + \frac{\pi^{3/2}}{4}
   p^{(3)}(\bfm{k},\bfm{k}') \cos(\theta_{\sbfm{k}} +
   \theta_{\sbfm{k}'} - \theta_{\sbfm{k} + \sbfm{k}'}),
\label{eq11}
\end{eqnarray}
where $\bfm{k} \ne \bfm{k}'$. These are the explicit forms of the
relation between phase correlations and the bispectrum in the
first-order approximation. We find that the distribution of the `phase
sum' $\theta_{\sbfm{k}} + \theta_{\sbfm{k}'} - \theta_{\sbfm{k} +
\sbfm{k}'}$ is determined only by the normalized bispectrum at the
first-order level, although higher-order normalized polyspectra can
contribute in general. The higher-order calculations show that the
distribution of the phase sum $\theta_{\sbfm{k}_1} + \cdots +
\theta_{\sbfm{k}_N}$ for the modes with closed wavevectors $\bfm{k}_1
+ \cdots + \bfm{k}_N = 0$ is determined by normalized polyspectra of
order 3 to $N$ in the lowest-order approximation, where the
identification $\theta_{\sbfm{k}} = - \theta_{-\sbfm{k}}$ is
understood. It was vaguely suggested that there is some relationship
between the phase sum and polyspectra based on a particular
non-Gaussian model by \citet{wat03a}. We now find the explicit
relationship between them in general non-Gaussian fields.

If we further integrate all phases but one particular
$\theta_{\sbfm{k}}$, the one-point probability function of a phase is
uniform, ${\cal P}(\theta_{\sbfm{k}}) = 1/2\pi$, which is consistent
with the previous $N$-body analysis \citep{sug91}. This conclusion
does not depend on the first-order approximation, since a single
wavevector can not be closed unless $\bfm{k} = 0$. Similarly, the
two-point probability function ${\cal P}(\theta_{\sbfm{k}},
\theta_{\sbfm{k}'})$ is also uniform unless $\bfm{k} = 2\bfm{k}'$. At
first glance, this conclusion seems to contradict the reported
non-uniform distribution of the phase difference of neighboring
wavevectors $D_{\sbfm{k}} \equiv \theta_{\sbfm{k}+\Delta\sbfm{k}} -
\theta_{\sbfm{k}}$ in $N$-body data
\citep{sch91,col00,chi01,chi02,wat03b}, where $\Delta\bfm{k}$ is a
fixed small vector. The same arguments are also applied to higher-order
approximations, so that the phase correlations between neighboring
wavenumbers should not appear even in strongly non-Gaussian fields in
a statistical sense.

To resove this puzzle, it is useful to consider the conditional
probability function given a Fourier coefficient of a small wavenumber
$\alpha_{\Delta\sbfm{k}}$. In the first-order approximation, the joint
probability of having phases $\theta_\sbfm{k},\theta_{\sbfm{k} +
\Delta\sbfm{k}}$ with fixed $\alpha_{\Delta\sbfm{k}}$ is given by
\begin{equation}
   {\cal P} \left( \theta_\sbfm{k},\theta_{\sbfm{k} +
   \Delta\sbfm{k}} \left|\alpha_{\Delta\sbfm{k}}\right. \right)
   \propto  1 + \frac{\pi}{2}
   |\alpha_{\Delta\sbfm{k}}| \cos\left( \theta_{\sbfm{k} +
   \Delta\sbfm{k}} - \theta_\sbfm{k} - \theta_{\Delta\sbfm{k}} \right)
   p^{(3)}(\Delta\bfm{k}, \bfm{k}),
\label{eq12}
\end{equation}
which arise the non-uniform distribution pattern of the phase
difference $D_\sbfm{k}$. The pattern depends on the fixed phase
$\theta_{\Delta\sbfm{k}}$, which means the pattern varies from sample
to sample, and this is exactly what is reported in the $N$-body
analyses. The pattern of the phase difference should be significant
for red power spectrum, which is also consistent with $N$-body
analyses. The functional form of equation (\ref{eq12}) also agrees
with the $N$-body analysis \citep{wat03b}. The statistics of phase
difference is thus the manifestation of the large-scale patterns of
individual realizations. The position of the trough in the
distribution of the phase difference corresponds to the phase of the
mode $\Delta \bfm{k}$, and the degree of deviations from the uniform
distribution depends on the specific amplitude of the mode
$\Delta\bfm{k}$ and also on the normalized bispectrum.

\section{A numerical demonstration}
\label{sec3}

The equations (\ref{eq10}) and (\ref{eq11}) relate the bispectrum to
the distribution of the phase sum $\theta_{\sbfm{k}} +
\theta_{\sbfm{k}'} - \theta_{\sbfm{k} + \sbfm{k}'}$. To see if this
kind of phase information is practically robust, we numerically
examine simple examples of non-Gaussian fields. Instead of examining
cosmological simulations, the following simple example is enough to
compare the numerical phase distributions and theoretical predictions.
Series of non-Gaussian fields are simply generated by exponential
mapping of a random Gaussian field:
\begin{equation}
   f(\bfm{x}) =
   \exp\left(g\phi(\bfm{x}) - g^2/2\right) - 1,
\label{eq13}
\end{equation}
where $\phi$ is a random Gaussian field with zero mean, unit variance,
and $g$ is the non-Gaussian parameter. We simply take a flat power
spectrum for the Gaussian field $\phi$. The field $f$ has zero mean
and variance $\langle f^2 \rangle = \exp(g^2) - 1$, and is called the
lognormal field \citep{col91}. This field has quite similar
statistical properties to gravitationally evolved non-Gaussian fields
and approximately follows the hierarchical model of higher-order
correlations. The parameter $g$ controls the non-Gaussianity, and the
random Gaussian field is recovered by taking the limit $g\rightarrow
0$. The random field $f$ is generated on $64^3$ grids in a rectangular
box with the periodic boundary condition.

\begin{figure}
\epsscale{0.6}
\plotone{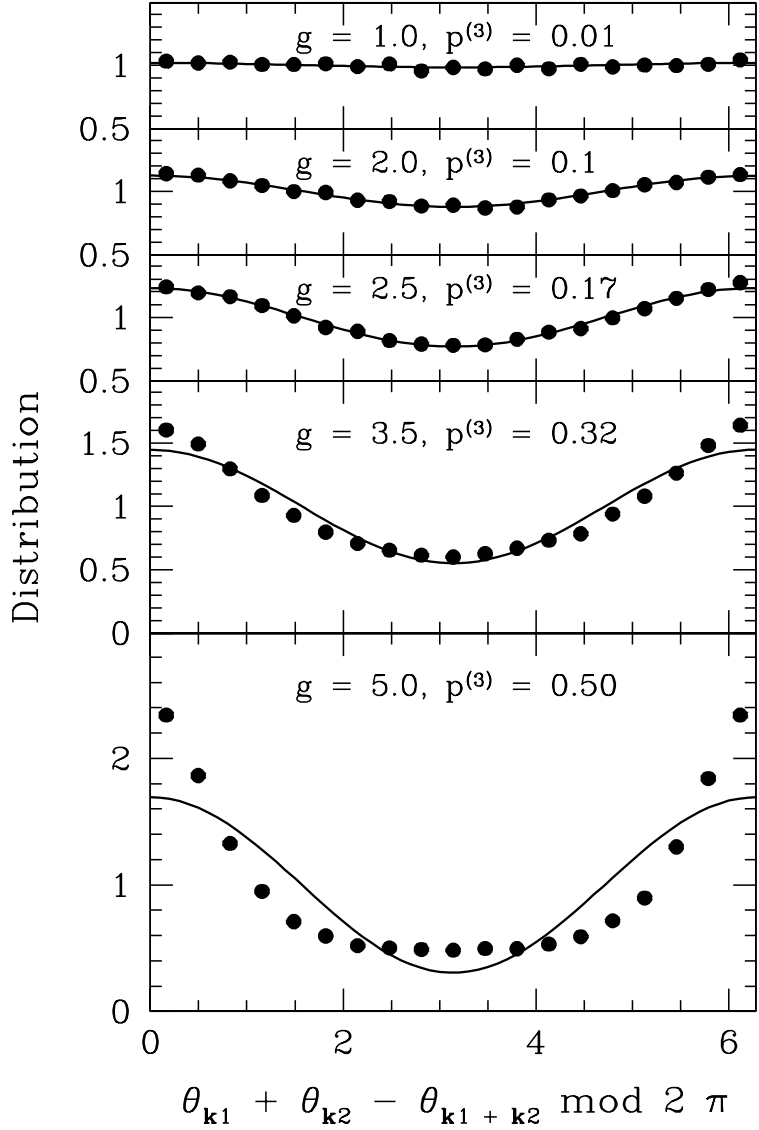}
\caption
{The distribution of the phase sum for a particular configuration of
wavevectors. Five non-Gaussian fields are shown, where $g$ is the
non-Gaussian parameter and $p^{(3)}$ is the normalized bispectrum for
the particular configuration. Theoretical predictions by normalized
bispectra in the first-order approximation are shown by solid
curves.
\label{fig1}}
\end{figure}

In Fig.~\ref{fig1}, the distribution of the phase sum
$\theta_{\sbfm{k}_1} + \theta_{\sbfm{k}_2} - \theta_{\sbfm{k}_1 +
\sbfm{k}_2}$ is plotted for a binned configuration of the wavevectors,
$|\bfm{k}_1| = [0.4,0.5]$, $|\bfm{k}_2| = [0.5,0.6]$, $\theta_{12} =
[50^\circ,60^\circ]$, as an example, where $\theta_{12}$ is the angle
between $\bfm{k}_1$ and $\bfm{k}_2$, and the magnitudes of the
wavenumber are in units of the Nyquist wavenumber.
The phase sum is averaged over the wavevectors in a configuration bin.
The points represent the distributions of the phase sum in each
realization. Poisson errorbars are smaller than the size of the
points. The normalized bispectra $p^{(3)}(\bfm{k}_1,\bfm{k}_2)$ are
numerically evaluated from each realization, which are used to draw
the theoretical curves in the first-order approximation of equation
(\ref{eq11}). There is not any fitting parameter at all. The agreement
is remarkable in weakly non-Gaussian fields. When the non-Gaussianity
becomes high, the data points deviate from the first-order
approximation, and the distribution of the phase sum is sharply peaked
at $\theta_{\sbfm{k}_1} + \theta_{\sbfm{k}_2} - \theta_{\sbfm{k}_1 +
\sbfm{k}_2} = 0\ {\rm mod}\ 2\pi$. Up to $g \sim 3.0$, or $\langle
f^2\rangle^{1/2} \simeq 100$, the distribution of the phase sum is
accurately described by the first-order approximation, and is
determined only by the normalized bispectrum. Even though the
non-Gaussianity $g \sim 3.0$ on scales of the Nyquist wavenumber is
beyond the perturbative regime, the normalized bispectrum on scales of
the presently tested configuration is still within the perturbative
regime, $p^{(3)} \sim 0.25$. This means that the phase sum is well
approximated by first-order formula of the present work even when the
field is strongly nonlinear in dynamics, as long as the parameter
$P(k)/V$ on the relevant scales is small. Increasing the power on
relevant scales and/or decreasing the volume drive the phase
correlation large, due to the fact that the phase correlations are
particularly dependent on significant features in the sample.

\section{Summary}
\label{sec4}

The structure of the phase correlations in non-Gaussian fields is
elucidated. The method to relate the joint distribution of phases to
polyspectra is newly found and developed. The distribution of the
phase sum of closed wavevectors is represented by the polyspectra. We
found the statistics of the phase difference reflect the particular
phase of the mode within an individual sample. The distribution of the
phase sum of three or more modes carries the statistically useful
information. The understanding of the phase correlations in
non-Gaussian fields is now reached unprecedented level in this Letter,
so that many investigations to make use of the phase information will
be followed, such as the analysis of the non-Gaussianity of all kinds
of cosmic fields, the nonlinear gravitational evolution of the density
fields, the biasing and redshift-space distortion effects on the
galaxy clustering, the primordial non-Gaussianity from inflationary
models, and so forth. One may also hope that phase information can be
useful in statistical analyses of all kinds of non-Gaussian fields,
from various phenomena of pattern formations to human brain mapping,
etc.

\acknowledgements

I acknowledge support from grants MEXT 13740150.

\clearpage


\clearpage

\end{document}